\documentstyle[11pt,paspconf,psfig,epsf]{article}
                                                 
\def \hi {H\,{\sc i~}}

\def \cmsq {$\rm cm\,^{-2}$}
\def \kmsws {$\rm km\,s^{-1}\,$}

\begin{document}
                
\title{HVCs probing a gaseous Galactic halo}
                                            
\author{Peter M.W. Kalberla, J\"urgen Kerp}
\affil{Radioastronomisches Institut der Universit\"at Bonn, Auf dem
H\"ugel 71, D-53121 Bonn, Germany}
                                  
\author{Urmas Haud}
\affil{Tartu Observatory, EE2444 Toravere, Estonia}
                                                   
\begin{abstract}
We study the hypothesis that high-velocity clouds (HVCs) may originate from
instabilities within
the gaseous phase of the Galactic halo.
According to the hydrostatic equilibrium model of Kalberla \& Kerp
(1998), we determine the
probability distribution of occurrence of instabilities within the
Galactic halo.
Instabilities may produce condensations within the Galactic halo
beyond a $z$-distances of 4\,kpc, which are accelerated by
gravity and decelerated by drag-forces.
                                           
We present evidence that the \hi high-velocity
dispersion component, discovered
by Kalberla et al. (1998) is associated with high-velocity-halo gas.
The physical properties of this high-velocity halo gas are similar to the
recently detected
highly-ionized HVCs by Sembach et al. (1995, 1998).
Accordingly, the \hi high-velocity dispersion component may be the neutral
tracer of the turbulent
gas motions within the Galactic halo.
Our calculations demonstrate, that the sky-averaged signal of in-falling
condensations does not differ
significantly from the \hi emission of the turbulent Galactic halo gas.
\end{abstract}
              
                                       
\section{Introduction}
                      
The origin of HVCs is still a matter of discussion.
Some of the HVCs seem to be located at extragalactic distances while
the large northern HVC complexes (M, A and C) are members of the Galactic halo
(Blitz et al. 1998, van Woerden et al. 1998).
                                             
In this paper we focus on the question: ``can high-velocity clouds
be explained as condensations of the Galactic halo matter?''.
For this aim, we study the stability of the gaseous Galactic halo and the
velocity distribution
of the in-falling condensations in the framework of a Galactic rain.
We include the drag-forces as well as the sweeping up of matter on their way
to the Galactic disk.
                     
Benjamin \& Danly (1997) investigated the influence of drag-forces on the
velocity of IVCs and HVCs.
Benjamin (these proceedings) argues that several observational facts
indicate interactions between HVCs and the interstellar medium:
(1) the existence of cometary shaped clouds in our Galaxy (Odenwald 1988),
(2) the correlation between distance and velocity of HVCs and IVCs claimed by
(Benjamin \& Danly 1997),
(3) the existence of velocity bridges or now called head-tail structures of HVCs
(Pietz et al. 1996),
and (4) the positional correlation between enhanced X-ray emission and HVCs
(Kerp et al. 1999).
                
Recent investigations of the \hi and X-ray data further supported the
HVC in-fall scenario:
Br\"uns (1998) searched for \hi head-tail structures of HVCs across the
whole sky which is covered
by the Leiden/Dwingeloo survey (Hartmann \& Burton 1997).
In total, he analyzed 252 HVCs with column densities
$N_{\rm \hi} > 10^{19}$\,\cmsq.
45 HVCs of his sample revealed head-tail structures.
Moreover, he deduced that the probability to find a HVC with a head-tail structure
increases proportional to the column density of the HVC.
                                                        
Kerp et al. (1999) searched for excess soft X-ray emission towards four
prominent HVC complexes.
Towards HVC complexes C, D and GCN/GCP, they detected excess soft X-ray emission.
In case of HVC complex C, they showed, that the excess soft X-ray emission
is in position closer
correlated with HVCs than with IVCs gas.
The majority of the X-ray bright HVCs have column densities
$N_{\rm \hi} > 5 \cdot 10^{19}$\,\cmsq.
Towards HVC complex C 2/3 of the head-tail structures studied by
Pietz et al. (1996) are associated
with excess soft X-ray emission.
                                
In this paper we study the development of neutral clouds falling
towards the Galactic disk.
These clouds are produced by instabilities within the Galactic halo gas.
The velocity and \hi brightness temperature distribution of these
condensations are quantitatively
compared with the Leiden/Dwingeloo data.
We present evidence, that low-column density neutral high-velocity gas
exists within the Galactic halo.
This high-velocity gas seems to be associated with the turbulent
motion of neutral gas condensations
in the halo.
Up to $|v_{\rm LSR}|\,\simeq\,350$\,\kmsws low-surface brightness
high-velocity gas is detectable in the new Leiden/Dwingeloo survey.
In Sect. 2 we present the basic parameters of a hydrostatic
equilibrium model of the Galaxy
according to the model of Kalberla \& Kerp (1998, hereafter K\&K ).
We address the stability of the Galactic halo and evaluate the
probability that individual neutral
condensations may be formed and fall onto the Galactic disk.
In Sect. 3 we compare the derived column density and velocity distribution of
our modeled HVCs with the observational data, and discuss the implications.
In Sect. 4. we discuss the implications of our results.
                                                       
\section{The model}
\subsection{The gaseous halo}
                             
Recently, K\&K showed that on large scales the Milky Way can be described well by a
hydrostatic equilibrium model.
K\&K included in their calculation the recent physical parameters of the
gaseous phases within the
Galactic halo.
They compared the model predictions with the observational situation using
the most recent
$\gamma$-ray, X-ray and radio surveys, which trace the distribution of
magnetic fields, cosmic-rays
and of the interstellar gas.
Their model consists mainly of 3 clearly distinct regions:
(1) the gaseous halo, hosting an ubiquitous X-ray emitting plasma and a
neutral component with
a high-velocity dispersion of 60\,\kmsws, in addition to the well know highly-ionized atomic species. (2) the disk, consisting predominantly of
cold- and warm-neutral medium with scale heights of 150 pc and 400 pc
respectively, and
(3) a disk-halo interface, which is the environment of the diffuse
ionized gas with a scale height
of 950 pc (Reynolds, 1997).
Such a layered disk-halo structure was found on average
to be stable against Parker instabilities.
K\&K pointed out, that the stability depends strongly on the composition
of the layers, the most
critical region is the disk-halo interface.
In any case, a hierarchical disk-halo composition is required for a stable halo.
                                                                                
One remarkable fact, which became apparent at this conference is, that
hydro-dynamical calculations by Avillez (1997 and these proceedings) resembles gas layers
with similar scale heights, densities and temperatures as deduced
by the hydrostatic equilibrium
model of K\&K.
Stability in a dynamical modeling requires a constant mass flow considering vertical upward
motions of a fountain flow and downward motion of the cooled gas.
                                                                 
This similarity may indicate, that the large scale view of the Milky Way
is indeed well approximated by a hydrostatic equilibrium model,
however, this does not imply that the halo is stable and in
equilibrium on all scales.
Stability requires that the gas pressure exceeds a minimum value
$p_{\rm gmin}(z)$:
\begin{equation}
p_{\rm gas}(z) > p_{\rm gmin}(z) = \frac{n^2(z) \; \partial \Phi/ \partial z}{
\partial n/ \partial z}.
\end{equation}
Here $n(z)$ is the gas density, $p_{\rm gas}(z)$ the gas pressure
and $\Phi(z)$ the gravitational potential perpendicular to the Galactic plane.
                                                                              
Instabilities may occur if the local pressure of the plasma in the halo
exceeds the steady state value significantly, e.g. if the plasma rises
to scale heights $h_z > 6 $ kpc. In such a case, the stability condition given in
Eq. 1 is violated.
The stability criterion in dynamical models implies, that next to the
rising gas some gas
condensations must fall back to the Galactic disk.
Here, we study a ``Galactic rain'' which is caused by instabilities
within the K\&K model beyond
a $z$-distance of 4\,kpc.
These instabilities form condensations within the highly 
turbulent gas phase of the Galactic halo.
                                                                                                  
\subsection{HVCs originating from local instabilities}
We assume, that the amount of halo gas which may
condense due to instabilities, is proportional to the local gas density
and proportional to the fractional overpressure which caused the instability.
In our approach, we assume that the production of a HVC may occur as a
stochastic process.
This neglects the spiral structure of the Milky Way,
and accordingly the
probability distribution of the perturbation across the Galactic disk.
Our approach is certainly a simplification of the real situation in detail, but on
large angular and long time scales, we will obtain statistical information on the
velocity distribution of the condensations.
                                           
Figure\,1 shows the probability distribution for the creation of an HVC
by instabilities.
Below about $|z|\,=\,3.8$\,kpc it is very unlikely that a condensation will
be formed out of the halo
material.
The probability reaches its maximum at $|z|\,=\,4.4$\,kpc, corresponding
to the average scale height of the gaseous halo. Beyond this $z$-distance
the probability decreases proportional to the volume density distribution.
                                                                          
%
\begin{figure}[th]
\centerline{
\psfig{figure=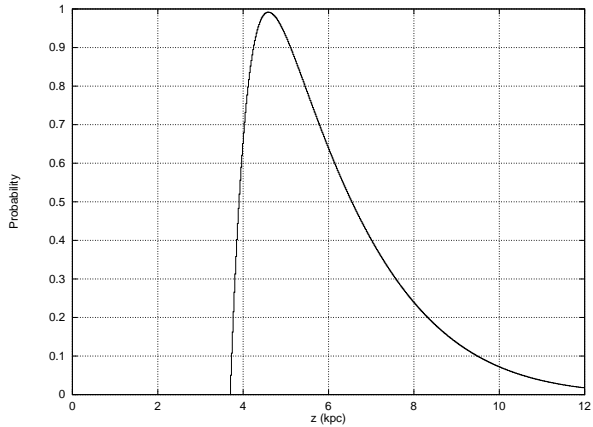,width=8cm,angle=-90}}
\caption[]{
In the framework of the hydrostatic equilibrium model of the
Galaxy (Kalberla \& Kerp 1998),
instabilities can evolve at high $z$-distances only.
According to this model, the relative probability to form condensations
within the Galactic halo
introduced by local instabilities reaches its maximum around the
$z$-distance of $\sim 4.4 $ kpc of the halo gas.
We expect that condensations which appear as HVCs originate
predominantly above $4 $ kpc.
\label{fig1} }
\end{figure}

\subsection{HVCs affected by gravity and drag }
                                               
Condensations from the gaseous halo are accelerated by gravity
until friction sets in,
which is caused by the layered structure of the Galaxy.
The further velocity development of the condensations and their final
appearance as an
intermediate-velocity cloud (IVC),
depends on the initial
column density of the condensation.
                                   
%
\begin{figure}[h]
\centerline{
\psfig{figure=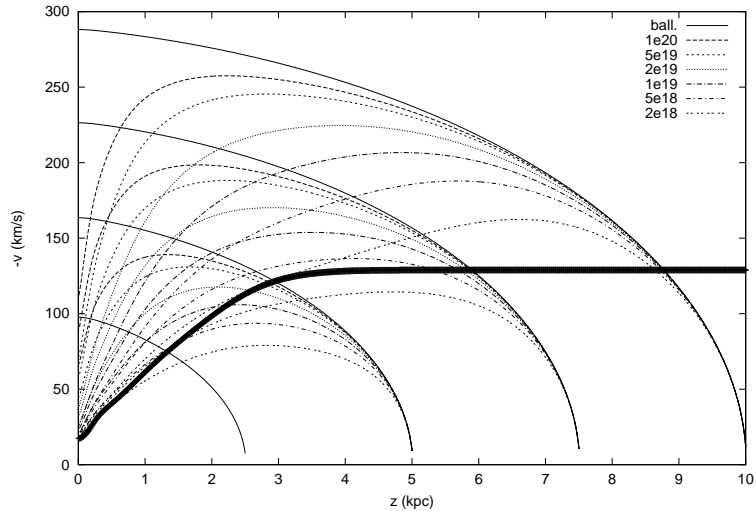,width=10cm,angle=-90}
}
\caption[]{
Velocities in $z$-direction for condensations originating in the
local vicinity of the Sun,
starting from $z$-heights of 2.5, 5, 7.5 and 10 kpc.
The clouds are accelerated by gravity and decelerated
by drag-forces according to our model assumptions. The solid lines represent
the ballistic case, which marks the in-fall of a condensation without
any deceleration.
Obviously, condensations originating from $z = 2.5 $ kpc are too slow to
be considered as HVCs.
For $z$-heights of  5, 7.5 and 10 kpc we plot
trajectories for HVCs with column densities of
$N_{\rm \hi} = 1 \cdot 10^{20}, 5 \cdot 10^{19}, 2 \cdot 10^{19}, 1 \cdot 10^{19},
5 \cdot 10^{18}, 2\cdot 10^{18}$\cmsq (top to bottom).
The thick line indicates the mean sound velocity within the Galactic halo.
Condensations above this line move supersonic, condensations below
move sub-sonic with respect to
the surrounding gaseous halo.
IVC velocities appear to be present at the beginning and end of the
in-fall of a condensation.
A low column density condensation high above the Galactic disk
will be certainly ionized by the
Galactic and extragalactic radiation field.
Thus, large $z$-distance IVCs are unlikely to be detectable in
\hi 21-cm line emission.
\label{fig2}}
\end{figure}

To evaluate the gravitational acceleration of the Milky Way, we adopt the equation
published by Kuijken \& Gilmore (1989).
The decelerating drag forces are parameterized according to Benjamin \& Danly (1997).
In addition, we introduce two major modifications.
First, to calculate the drag forces we use the gas densities as determined by K\&K.
Second, we assume that the condensations are sweeping up gas as they approach the
Galactic disk.
              
HVCs may either sweep up material on their way through the halo or
they may loose gas
which is stripped off from the outer cloud boundaries due to drag forces. Probably,
both effects occur at the same time, however we assume that on the
average more material
is swept up by the HVCs than lost.
The amount of matter which a HVC is sweeping up is highly uncertain.
We estimate that on the average HVCs accumulate 50\% of the gas which is passed by.
Such a rate seems to be reasonable because IVCs have
significantly higher column densities
than HVCs. In addition at such a rate we obtain in our model
calculations IVC velocities which are close to the observed ones.
As a consequence of gas enrichment,
in our model the gas-to-dust ratio as well as the
metalicity of HVCs and IVCs is modified by the swept-up material.
                                                                 
Figure 2 shows the velocity distribution of a sample of condensations
with different column densities.
We calculate trajectories for HVCs originating at $z$ = 5, 7.5 and 10 kpc.
Gravity and drag forces are evaluated in the solar vicinity.
In each case the $z$-velocities are given for column densities between
$N_{\rm \hi} = 2 \cdot 10^{18}\,{\rm cm^{-2}}$ and
$N_{\rm \hi} = 1 \cdot 10^{20}\,{\rm cm^{-2}}$,
and for comparison the ballistic curve without any energy loss.
For $|z| = 2.5 $ kpc only the ballistic curve is given.
Obviously, condensations with low column densities are significantly decelerated
by drag forces.
The higher the column density of the condensation, the higher the maximum speed.
Condensations with high column densities reach their maximum
velocities within the $z$-distance
range of $ 1 < |z| < 3 $ kpc,
while clouds with low column densities have their maximum velocities at
large $z$-distances.
Close to the Galactic disk ($|z| < 400 $ pc) our model predicts condensations
with IVC velocities in the range $20 < v_{\rm z} < 100$ \kmsws.
Our main conclusion from Fig. 2 is, that condensations which are formed at
distances $|z| > 4 $ kpc may appear as HVCs.
                                            
According to the hydrostatic equilibrium model of K\&K, we can compare
the velocity of the
condensations with the sound velocity (bold solid line in Fig. 2) in
the Galactic halo.
At $|z| > 4 $ kpc $v_{\rm s}(z) = 130 \; \rm km\,s^{-1}$ while
close to the disk $v_{\rm s}(z)$ drops to 25 $\rm km\,s^{-1}$.
Condensations which have velocities above this line are supersonic,
those below move sub-sonic.
For most of the $z$-distances, condensations with
$N_{\rm \hi} > 10^{19}$ cm$^{-2}$ move supersonic with
respect to the gaseous halo.
                            
\subsection{The Galactic rain}
                              
Up to now we have shown that above $|z|\,\geq\,4$\,kpc perturbations
in the Galactic halo gas may cause instabilities.
We demonstrated, that the condensations caused by such instabilities
may reach velocities comparable to the observed
HVC velocities, otherwise they might appear as IVCs.
                                                    
Now, we introduce more quantitative aspects into the discussion.
First, to overcome the arbitrary boundary condition that the
analysis is restricted to the local
neighborhood of the Sun, we extend the calculation to the entire Milky Way.
We use the density distribution according to K\&K.
Second, we introduce a random motion of the halo gas into our model,
which is indicated by the
detection of the \hi high-velocity dispersion component by
Kalberla et al. (1998).
Third, we include a co-rotation of the Galactic halo with the disk,
according to the result of Savage et al. (1997). The rotation curve is taken from
Fich et al. (1990).
Finally, the column density distribution of the condensation should
resemble the observed
column density distribution of the HVCs (Murphy et al. 1995).
                                                             
These assumptions allow to construct a model which is consistent with the known physical conditions
within the Galactic halo.
According to this model, we now generate a ``Galactic rain''
falling down
onto the Milky Way, triggered by random events across the entire halo.
The rain falls during a time twice the free-fall time of the lowest column
density condensations. After this period, we stop the calculation and evaluate
the ``frozen-in'' column density and velocity distribution of the condensations.
In the final step, we quantitatively compare the velocity as well as the
column density
distribution of the model with observational \hi data.
                                                      
\section{Comparison between model and observations }

%
\begin{figure}[ht]
\centerline{
\psfig{figure=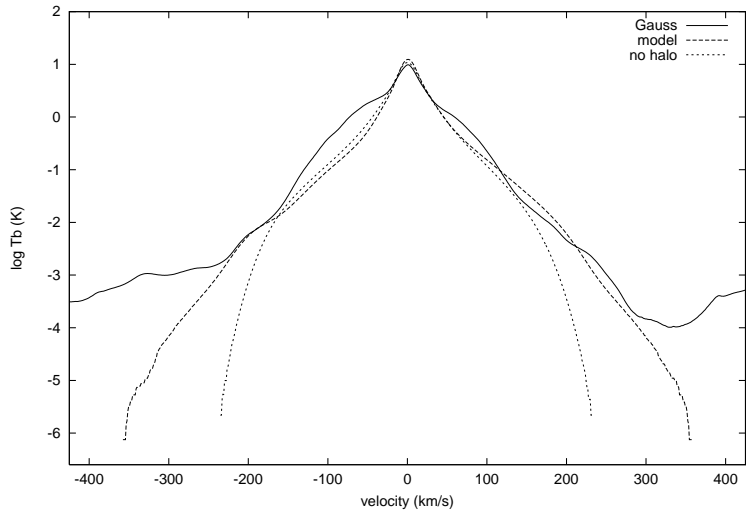,width=10cm,angle=-90}}
\caption[]{
The brightness temperature $T_{\rm b}$ distribution averaged 
across the entire Galactic sky covered
by the Leiden/Dwingeloo data is plotted versus $v_{\rm LSR}$.
The solid line represents the observational data, decomposed into Gaussian components.
The long dashed line marks the Kalberla \& Kerp (1998) hydrostatic equilibrium model.
Most important is, that the hydrostatic equilibrium model fits the data quantitatively well up to the
high-velocity gas regime.
The velocity dispersion of the Galactic disk and the disk-halo
interface cannot account for the detected \hi emission in the high and very-high velocity regime
(dotted line).
Accordingly, the hydrostatic equilibrium model of Kalberla \& Kerp (1998) predicts the faint
high-velocity \hi emission caused by turbulence within the Galactic halo.
\label{fig3} }
\end{figure}
                                      
Our approach is optimized to derive overall statistical properties of
in-falling condensations onto the Milky Way.
We have to compare the modeled situation with the observed one.
In Fig.\,3 we plot the mean observed brightness temperature $T_{\rm b}$
on a logarithmic scale versus the radial velocity ($v_{\rm LSR}$).
The solid line in Fig. 3 shows the Leiden/Dwingeloo \hi data
decomposed into Gaussian components.
All components with a significance of at least 3 $\sigma$ have been
integrated. The main reason using Gaussian components is
the suppression of the rms-noise at high velocities.
For comparison with the observation we plot the \hi distribution
according to the K\&K model (dashed line). For positive velocities $v_{\rm LSR} < 300$
\kmsws model and observations agree well. For negative velocities
in the range $-180 < v_{\rm LSR} < -20 $ \kmsws excess emission is observed,
which is associated with the inner part of the Galaxy beyond the scope of the K\&K model.
The excess emission for $v_{\rm LSR} < -220 $ \kmsws
was found to be
predominantly due to the Magellanic Stream and the Anti-Centre-Complex.
The dotted line in Fig. 3 represents a simulation, derived from the K\&K model
{\em without} any gas in the Galactic halo.
Comparing the models with the observational data, it is
obvious, that the main \hi emission at velocities $|v_{\rm LSR}| > 200 $ \kmsws
is dominated by the turbulent neutral Galactic halo gas.
                                                        
In Fig. 3 the observed brightness distribution is biased due to the fact
that the Leiden/Dwingeloo survey covers only declinations $\delta > -30
\deg$. In Fig. 4 only observational data for latitudes
$b > 0 \deg$ are compared with the model.
Comparing the observations (solid
line) with the modeled \hi distribution (dashed line) we find, within
the uncertainties, an agreement in the velocity range
$ 160 < |v_{\rm LSR}| < 350$ \kmsws.
The velocity regions $ -300 < v_{\rm LSR} < -200 $ and $ v_{\rm LSR} > 300 $ \kmsws are
affected by residual baseline uncertainties in addition 
to those discussed by Kalberla et al. (1998).
The deviations between both curves in this range are probably due to instrumental uncertainties.
For the northern Galactic hemisphere we find no indications for
significant amounts of HVC gas which deviate from a distribution
predicted by the K\&K model. Thus, within our limited global
investigations the only HVCs which were found to be obviously incompatible
with a Galactic rain model are the Magellanic Stream and
Anti-Centre-Complex.
                    
%
\begin{figure}[th]
\centerline{
\psfig{figure=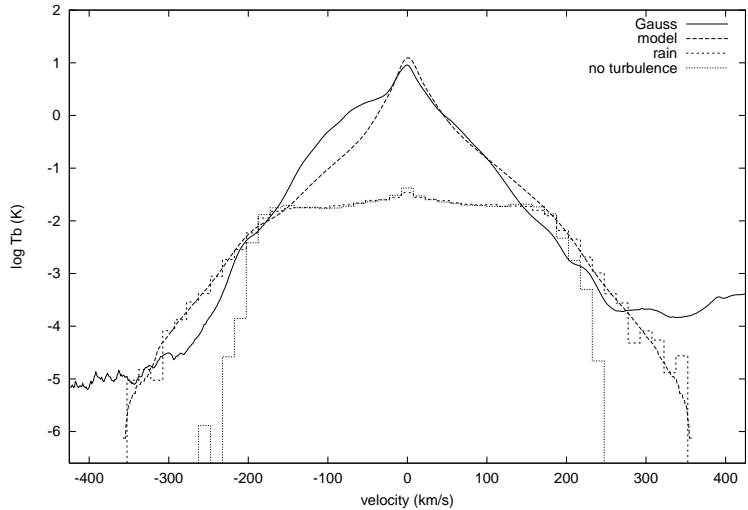,width=10cm,angle=-90}}
\caption[]{
The \hi brightness temperature distribution across the northern Galactic sky ($b\,>\,0\deg$) is
plotted versus $v_{\rm LSR}$.
The solid line marks the observational data decomposed into Gaussian components.
The long dashed line indicates the Kalberla \& Kerp (1998) hydrostatic equilibrium model.
The short dashed line shows the ``Galactic rain'' model.
Both models are the same for velocities with $|v_{\rm LSR}|\,>\,150$ \kmsws.
This is because the turbulent motion within the Galactic halo gas determines the motion of the
in-falling condensations until strong friction starts to dominate
close to the disk-halo interface region.
If we neglect the turbulent motion of the Galactic halo gas, as initial velocity of each
condensation, the velocity dispersion is much weaker (dotted line).
In particular, no very-high velocity \hi emission is predicted.
\label{fig4} }
\end{figure}

We can conclude, as an intermediate result, that most of the Galactic \hi emission at high
velocities $ |v_{\rm LSR}| > 250 $ \kmsws is caused by the turbulent 
neutral Galactic
halo gas. Now, we evaluate the column density distribution of the in-falling
condensations as described in Sect. 2.
The histogram marked by the dashed line in Fig. 4 represents the derived brightness temperature
distribution. At velocities $ |v_{\rm LSR}| > 180 $ \kmsws the
modeled HVC distribution is closely related to the model \hi
distribution derived for a gaseous halo.
Accordingly, considering {\em mean} properties only, we cannot
distinguish between the \hi emission of the in-falling condensations and
the \hi emission of the turbulent Galactic halo gas.
                                                    
Dropping the assumption that the condensations have an
initial velocity according to the
turbulent gas motion within the Galactic halo, we are able to
separate the signal of the in-falling condensations
from that of the turbulent Galactic halo gas.
The corresponding brightness temperature distribution of the ``non-turbulent'' condensations is
plotted as a histogram marked by the dotted line. Condensations originating form a non-turbulent
medium barely reach velocities exceeding $|v_{\rm LSR}| > 200 $ \kmsws.
In this case the derived velocity distribution resembles that of a
Galactic fountain (Bregman 1980, Wakker 1990).
Our conclusion is, that the turbulent Galactic halo gas contributes significantly to the very-high-velocity and
high-velocity \hi emission across the entire Galactic sky.

\section{Summary and conclusions}
In this paper we investigate the hypothesis that HVCs may originate
from instabilities within the Galactic halo.
Using the hydrostatic model by K\&K, we predict
the vertical distribution of \hi condensations originating
from such instabilities. HVCs originate predominantly above
$z$-distances of $\sim$ 4 kpc.
Considering gravitational acceleration and
deceleration by drag forces (Benjamin \& Danly 1997) we calculate
trajectories for such clouds and model their large scale velocity
distribution.
             
The velocity of an individual condensation depends on the initial mass.
The higher the $z$-distance and mass of the condensation, the
higher the terminal velocity. Strong deceleration starts when a
condensation approaches the Galactic disk-halo interface. Most
of the HVCs with column densities exceeding
$N_{\rm \hi} > 10^{19}$ cm$^{-2}$ move supersonic with
respect to the gaseous halo. For these clouds indications for
interactions with the interstellar medium are found:
head-tail structures and excess soft X-ray emission.
Further we find that the fraction of supersonic HVCs increases
proportional with column density.
                                 
Down to the very limits of the Leiden/Dwingeloo \hi data the hydrostatic
equilibrium model of K\&K fits the data well, across 6 orders of magnitude.
Averaged across the entire Galactic sky, the \hi lines of the
in-falling condensations are distributed similar to that of
the \hi 21-cm line emission of the turbulent Galactic halo gas.
In particular, assuming that HVCs originate from a
turbulent Galactic halo gas, represents the observed very-high-velocity
gas up to $|v_{\rm LSR}| < 350 $ \kmsws better than
Galactic fountain models.
                         
We conclude, that the turbulent Galactic halo gas produces
faint high-velocity \hi 21-cm line emission detectable across the
entire Galactic sky. Within this pervasive gas phase
condensations can be observed
as HVCs. On average, the velocity distribution
of HVCs does not deviate significantly from the distribution of the
pervasive \hi halo gas phase. HVCs share the turbulent motions within the
halo and are significantly affected by drag forces.
                                                   
The physical conditions within the turbulent Galactic halo gas are
comparable to those of the
highly-ionized HVC discovered by Sembach et al.
(1995, 1998 and these proceedings).
According to the K\&K model, the volume density at a $z$-distance
of 15 kpc is in the oder of $n_{\rm H}\,=\,10^{-6}\,{\rm cm^{-3}}$.
The pressure of the halo gas at such a $z$-distance is
$P$/k$\sim\,2\,{\rm cm^{-3}\,K}$, assuming a plasma
temperature of Galactic halo gas of
$T_{\rm plasma}\,=\,1.5\,10^6$\,K (Pietz et al. 1998).
We conclude, that the C{\sc iv} clouds discussed by Sembach et al.
(these proceedings) may be located within the outer Galactic halo.
Due to the intergalactic radiation field, only little \hi gas is
expected to be associated with HVCs at such distances.
                                                      

\end{document}